\documentclass[sigconf, screen=true,]{acmart}

\AtBeginDocument{%
  \providecommand\BibTeX{{%
    \normalfont B\kern-0.5em{\scshape i\kern-0.25em b}\kern-0.8em\TeX}}}

\setcopyright{acmcopyright}
\acmPrice{15.00}
\acmDOI{10.1145/3368089.3417046}
\acmYear{2020}
\copyrightyear{2020}
\acmSubmissionID{fse20ind-p5-p}
\acmISBN{978-1-4503-7043-1/20/11}
\acmConference[ESEC/FSE '20]{Proceedings of the 28th ACM Joint European Software Engineering Conference and Symposium on the Foundations of Software Engineering}{November 8--13, 2020}{Virtual Event, USA}
\acmBooktitle{Proceedings of the 28th ACM Joint European Software Engineering Conference and Symposium on the Foundations of Software Engineering (ESEC/FSE '20), November 8--13, 2020, Virtual Event, USA}

\usepackage{todonotes}
\usepackage{algorithmic}
\usepackage{algorithm}
\usepackage{tcolorbox}
\usepackage{paralist}
\usepackage{multirow}
\usepackage{booktabs}
\usepackage{graphicx}
\usepackage{float}
\usepackage{svg}
\usepackage{rotating}
\usepackage{array}
\usepackage{microtype}

\begin{document}

\title{Can Microtask Programming Work in Industry?}

\author{Shinobu Saito}
\email{shinobu.saitou.cm@hco.ntt.co.jp}
\affiliation{%
  \institution {Software Innovation Center, NTT Corporation}
  \city{Tokyo}
  \country{Japan}
 }
 
  \vspace{15mm}

 \author{Yukako Iimura}
\email{yukako.iimura.vr@hco.ntt.co.jp}
\affiliation{%
  \institution {Software Innovation Center, NTT Corporation}
  \city{Tokyo}
  \country{Japan}
  }

\author{Emad Aghayi}
\email{eaghayi@gmu.edu}
\affiliation{%
  \institution{George Mason University}
  \streetaddress{4400 University Drive}
  \city{Fairfax}
  \state{Virginia}
  \postcode{22030}
  \country{USA}
  }

 \author{Thomas D. LaToza}
\email{tlatoza@gmu.edu} 
\affiliation{%
  \institution{George Mason University}
  \streetaddress{4400 University Drive}
  \city{Fairfax}
  \state{Virginia}
  \postcode{22030}
  \country{USA}
  }

\begin{abstract}
A critical issue in software development projects in IT service companies is finding the right people at the right time. By enabling assignments of tasks to people to be more fluid, the use of crowdsourcing approaches within a company offers a potential solution to this challenge. Inside a company, as multiple system development projects are ongoing separately, developers with slack time on one project might use this time to contribute to other projects. In this paper, we report on a case study of the application of crowdsourcing within an industrial web application system development project in a large telecommunications company. Developers worked with system specifications which were organized into a set of microtasks, offering a set of short and self-contained descriptions. When crowd workers in other projects had slack time, they fetched and completed microtasks. Our results offer initial evidence for the potential value of microtask programming in increasing the fluidity of team assignments within a company. Crowd contributors to the project were able to onboard and contribute to a new project in less than 2 hours. After onboarding, the crowd workers were together able to successfully implement a small program which contained only a small number of defects. Interview and survey data gathered from project participants revealed that crowd workers reported that they perceived onboarding costs to be reduced and did not experience issues with the reduced face to face communication, but experienced challenges with motivation.
\end{abstract}




\maketitle

\section{Introduction}

\begin{figure}
\centering
\includegraphics[width=\columnwidth,keepaspectratio, clip]{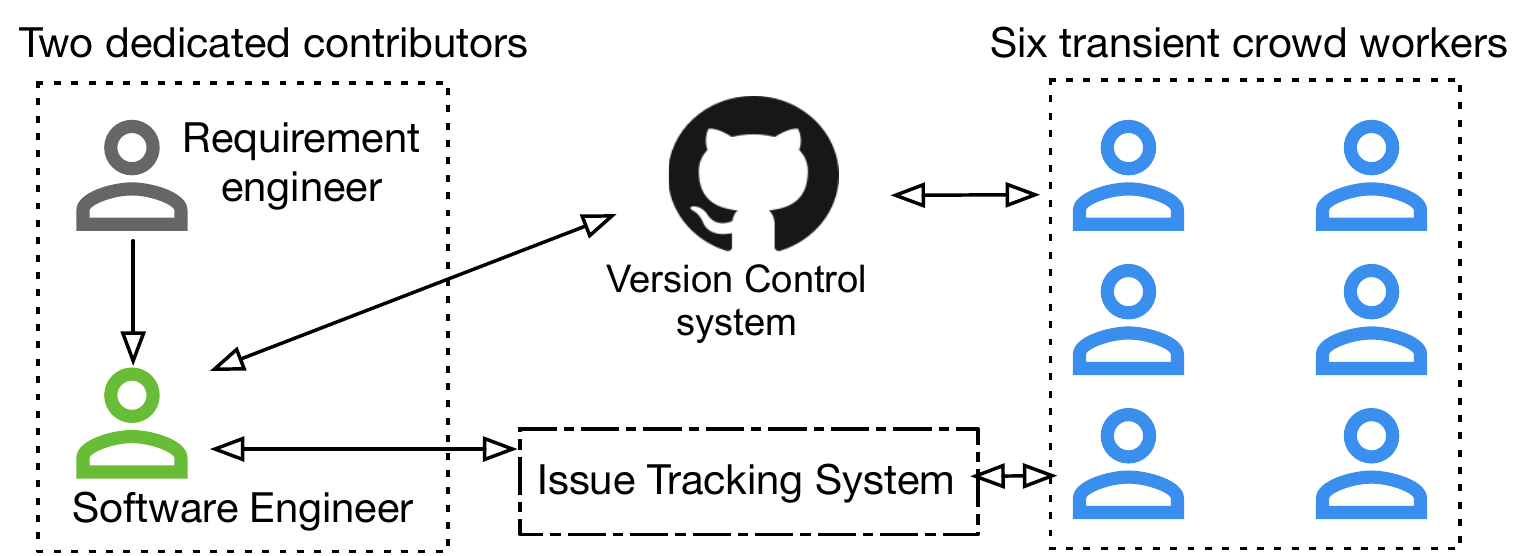}

\caption{An overview of the project organization. Two dedicated full-time engineers assigned to the project managed and organized the project, and six crowd workers assigned to other projects within the company contributed in their slack time. 
}
\label{fig:members}
\end{figure}

A critical issue in software development projects in IT service companies is finding developers with the right knowledge. For example, one report suggests there will be a shortage of 430,000 IT specialists in 2025 in Japan~\cite{Cliff2025}. When projects require more resources than available, it is often challenging to quickly meet these needs. Recruiting new employees from outside a company requires a substantial investment of time and cost. Developers might instead be brought in from other ongoing system development projects or teams inside the company. However, even for developers already working within a company, it can require a substantial investment of time and effort to successfully onboard them onto a project. For example, developers need to learn project background knowledge such as the system architecture, project configuration, and coding conventions. For these reasons, balancing resources within a company often does not work effectively~\cite{aurum2013a,j2003a,bj2008a,fagerholm2014a}. \par

One potential solution to these challenges is the use of crowdsourcing~\cite{d2018a,latoza2015a,latoza2014a,aghayi2019implementing}. In crowdsourced software engineering, work traditionally done inside a company is outsourced to an undefined crowd in the form of an open call~\cite{mao2017a}. Successful commercial platforms for crowdsourcing software engineering work include TopCoder, AppStori, uTest, and TestFlight. Crowdsourcing offers companies the potential for fluidity, enabling workers from outside to be recruited on demand at the moment in which work must be completed. One form of crowdsourcing is microtask programming, where programming tasks are decomposed into short, self-contained tasks with a clear objective~\cite{latoza2015a}. By decomposing the specifications of a system into small, self-contained descriptions, microtasking aims to further decontextualize work, enabling even greater fluidity by reducing the needed context to onboard. \par

While crowdsourcing, by definition, involves recruiting contributors from outside a company or organization, it might also be possible to apply crowdsourcing within a company. Rather than offer work to external developers, developers working on other projects within a company with available slack time might use this time to complete microtasks and contribute to another project. For companies with closed-source code and confidential information and intellectual property to protect, this model offers many of the potential crowdsourcing benefits of lower onboarding costs and greater resource fluidity with fewer of the potential drawbacks. \par

In this paper, we report on a project at NTT which applied microtask programming to an industrial web application development project. Inspired by behavior-driven microtask programming~\cite{aghayi2019implementing}, a workflow was used where crowd workers could make two types of microtask contributions: 1) implement a micro-specification or 2) review an implemented micro-specification. Crowd workers followed Test Driven Development (TDD) to implement micro-specifications. Micro-specifications were defined and collected in a ticket pool, and developers with slack time on their own project were able to fetch and complete microtasks. Inspired by the role of the co-pilot in TopCoder~\cite{stol2014two} responsible for managing and overseeing development work, the work of decomposing and integrating microtasks was done by two engineers dedicated to the project while six internal crowd workers completed microtasks ( Fig.~\ref{fig:members}). The dedicated software engineer served three roles in the project, working as a software designer to create microtasks, as a developer to implement several tasks, and as a tester.\par

Overall, the project achieved its objectives and a web application for managing finance closing processes was successfully created. Crowd workers together implemented approximately 2800 LOC. All were able to initially onboard onto the project in less than 2 hours and successfully communicated via an issue tracking system (ITS). Crowd workers reported that they perceived onboarding costs to be reduced and did not experience issues with the reduced face-to-face communication, but did experience challenges with motivation.
\par

The remainder of this paper is organized as follows. Section 2 describes related work in crowdsourced software development, microtask programming, and inner sourcing. Section 3 describes the microtask programming process and tools adopted in the case study. Section 4 first presents results on the outcome of the project, including onboarding activities, and the describes the perceptions of the project participants about the use of a microtask programming process inside a company. Section V discusses the implications of our findings, and Section VI concludes.\par

\section{Related Work}
A variety of work has explored crowdsourcing approaches for software development activities such as design~\cite{t2015a}, implementation~\cite{d2018a,aghayi2019implementing,chen2017a,s2015a}, and testing~\cite{dwarakanath2015a}. Companies such as TopCoder, uTest, and UserTesting.com offer a platform and community for crowdsourcing software activities to an external crowd. A study of TopCoder found a number of factors that impact project quality, including the number of contemporary projects, the length of documents, and the number of registered developers~\cite{stol2014two}. \par

When developers join a new project, they may face a number of onboarding barriers, including installing necessary tools, identifying and downloading dependencies, and configuring their build environment, understanding the codebase, and identifying a task~\cite{joiningOSS_Alex2003,steinmacher2015a,jergensen2011a}. As a result, successfully onboarding onto a new project may require weeks of time, creating a substantial barrier to fluidly assigning developers to match project resource needs.\par

Crowdsourced programming environments have been designed to reduce some of these barriers by offering a preconfigured programming environment. In Apparition, developers create microtasks for crowd workers to build small user interface elements and their behaviors, which crowd workers can then complete in a dedicated environment~\cite{s2015a}. In CodeOn, developers speak requests for small contributions which other developers can then supply~\cite{chen2017a}. In microtask programming, developers complete short microtasks in a dedicated environment in which they are given a function or test and are asked to make a small change to it~\cite{aghayi2019implementing,d2018a,MicrotaskProgrammingGC}. By decontextualizing work from the larger project context and offering a dedicated environment with the necessary background information and editors to make the contribution, developers are able to make contributions in under 5 minutes with only a short onboarding period. Studies of the use of crowdsourcing in industry have found that companies have relatively low awareness of these new approaches ~\cite{prikladnicki2014a}.\par

A key challenge in crowdsourced software development is in designing effective coordination mechanisms, which may have a variety of benefits and drawbacks~\cite{goldman2012a,goldman2011a}. One approach is for coordination between the requirements engineer and crowd workers to occur entirely through the specifications themselves~\cite{aghayi2019implementing}. An alternative approach is for the requirements engineer or developer responsible for creating the microtasks to directly coordinate with each worker~\cite{s2015a}. Another key choice is the ways in which crowd workers themselves may coordinate, such as through instant messaging~\cite{aghayi2019implementing} or through more structured interactions such as around issues~\cite{latoza2015b}. Crowd workers themselves have expressed a desire for direct worker-to-worker communication when handoffs or reviews necessitate interactions between workers~\cite{latoza2015b}.\par

Noting the success of open source approaches, companies have explored adopting open source software development practices inside their organizations through a set of practices named inner sourcing~\cite{m2016a}. Developers working using inner sourcing do not belong to a single project, and anybody in the organization may contribute to all projects in it~\cite{k2015a}. For companies building closed-source software with confidential information and intellectual property, inner source offers new techniques and models for encouraging internal collaboration by applying open-source best practices within organizations~\cite{unknown-a,unknown-b}. Studies examining the impact of inner sourcing found potential benefits such as increased development efficiency, higher code quality, and quicker development cycles~\cite{m2016a}.\par

Our work builds on these existing techniques and studies, focusing on the potential fluidity offered by adopting microtasked development in an industrial context. 
\begin{figure}[h]
\centering
\includegraphics[width=\columnwidth,keepaspectratio, clip]{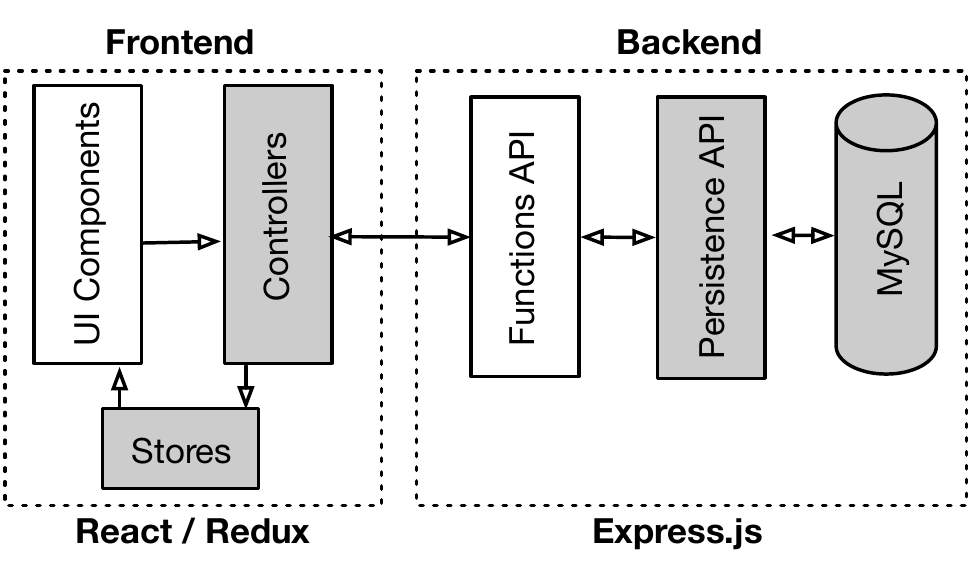}

\caption{The project consisted of a frontend and backend with 6 major components. In both the frontend and backend, components which required more project knowledge were implemented by the dedicated software engineer (gray background) while those which required less project knowledge were implemented by the crowd workers (white background).}
  \label{fig:architect}
\end{figure}

\begin{figure*}[ht]
\centering
\includegraphics[width=\textwidth,keepaspectratio, clip]{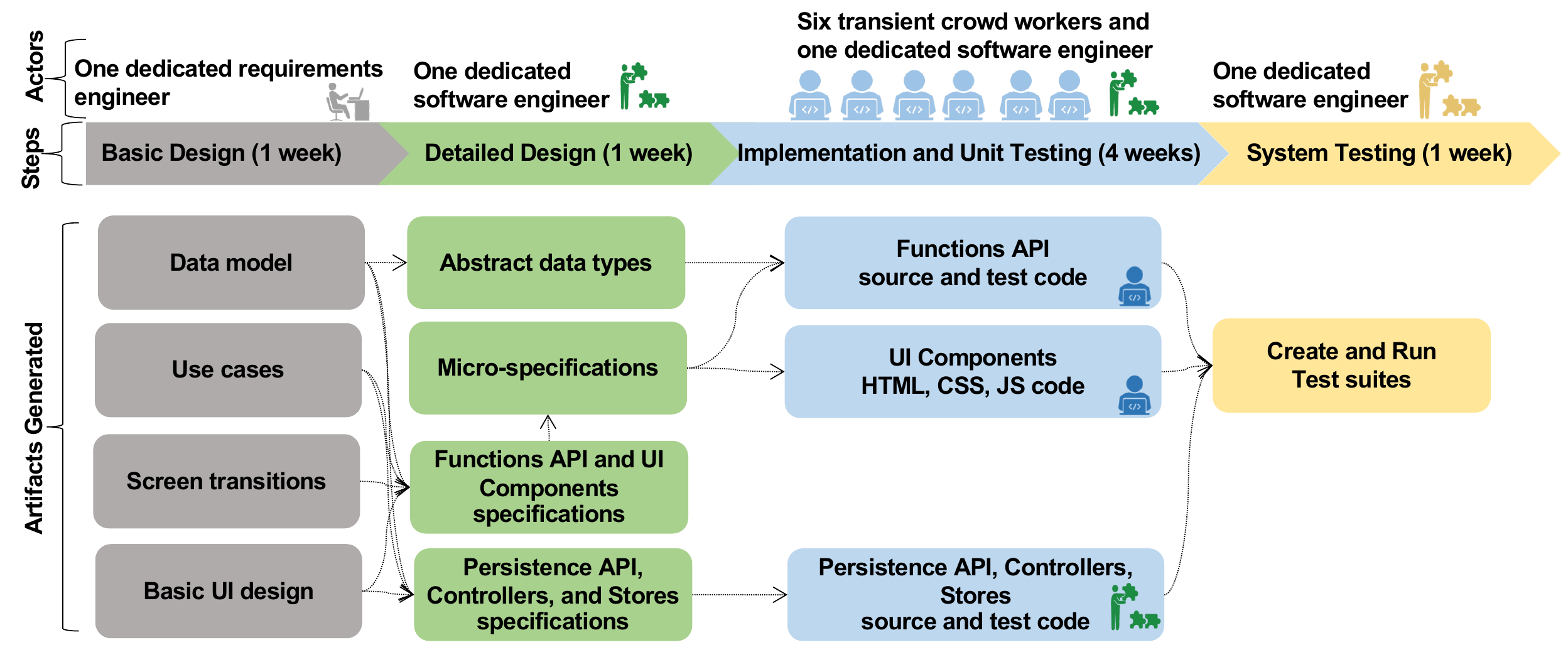}

\caption{Microtask programming was applied using a four step workflow. A requirements engineer first designed a data model, use cases, and user interface. A dedicated software engineer then created specifications and microtasks. Six transient crowd workers then completed the microtasks in their available slack time. At the same time, the dedicated software engineer implemented components that required more project knowledge. Finally, the dedicated engineer created and executed a test suite. After seven weeks, the project was designed, implemented, and tested.} 
\label{fig:GeneralOverview}
\end{figure*}

\section{Microtask Programming in a Company}
\subsection{Project Overview}
In the project, an in-house web application system for managing information on finance closing processes was built. The existing system had been used by an organization within NTT, had about 50 users, and had operated for five years. Due to business reasons including the expiration of a license, it was decided to reimplement the system. In order to more fluidly assign resources to the project, the project was organized using a crowdsourced software development approach.

\subsection{Organization Structure}
All participants in the project were employees of NTT. As shown in Figure~\ref{fig:members}, two dedicated engineers were assigned to work full time on the project, a requirements engineer and a software engineer. In addition, 6 crowd workers, who were primarily assigned to other projects inside the company, were asked to make use of their available slack time to contribute to the project. For example, if they was not busy at the end of the day or in the morning, they might use this time to contribute. The dedicated engineers and crowd workers were located in geographically distributed areas, with the dedicated engineers located in a single site and the crowd workers located at a different site.

\subsection{System Architecture}
Parts that required more project knowledge were assigned to the dedicated software engineer, while those which required less were implemented by the crowd workers (Figure~\ref{fig:architect}). The crowd workers implemented two parts of the web application (UI Components and Functions API), and the dedicated software engineer implemented three (Controllers, Stores, and Persistence API). Implementing these three required more knowledge of the data model and its complexities. The web application was implemented using React and Redux in the frontend and Express.js and MySQL in the backend. \par

In the frontend, the dedicated software engineer implemented the Controllers and Stores, and crowd workers implemented the UI Components. The UI Components were first decomposed into micro-specifications by the dedicated software engineer, and the crowd workers then implemented them. The micro-specifications in the frontend did not have mutual dependencies, and there was not a specific order in which they needed to be completed.  \par

In the backend, the dedicated software engineer implemented the Persistence API, and the crowd workers implemented the Functions API. The Persistence and Functions API did not have dependencies. The dedicated software engineer generated several micro-specifications for the Functions API, which had mutual dependencies. This required managing the order in which they were implemented.\par

\subsection{Software Development Process}
Figure \ref{fig:GeneralOverview} depicts the overall development process used and the artifacts generated at each step. The development process consisted of four steps: basic design, detailed design, implementation and unit testing, and system testing. Basic design was completed by the requirements engineer in the traditional way. The requirements engineer created four artifacts: a conceptual data model, use cases, screen transitions, and a basic UI design. \par

\begin{figure}
\centering
\includegraphics[width=\columnwidth,keepaspectratio, clip]{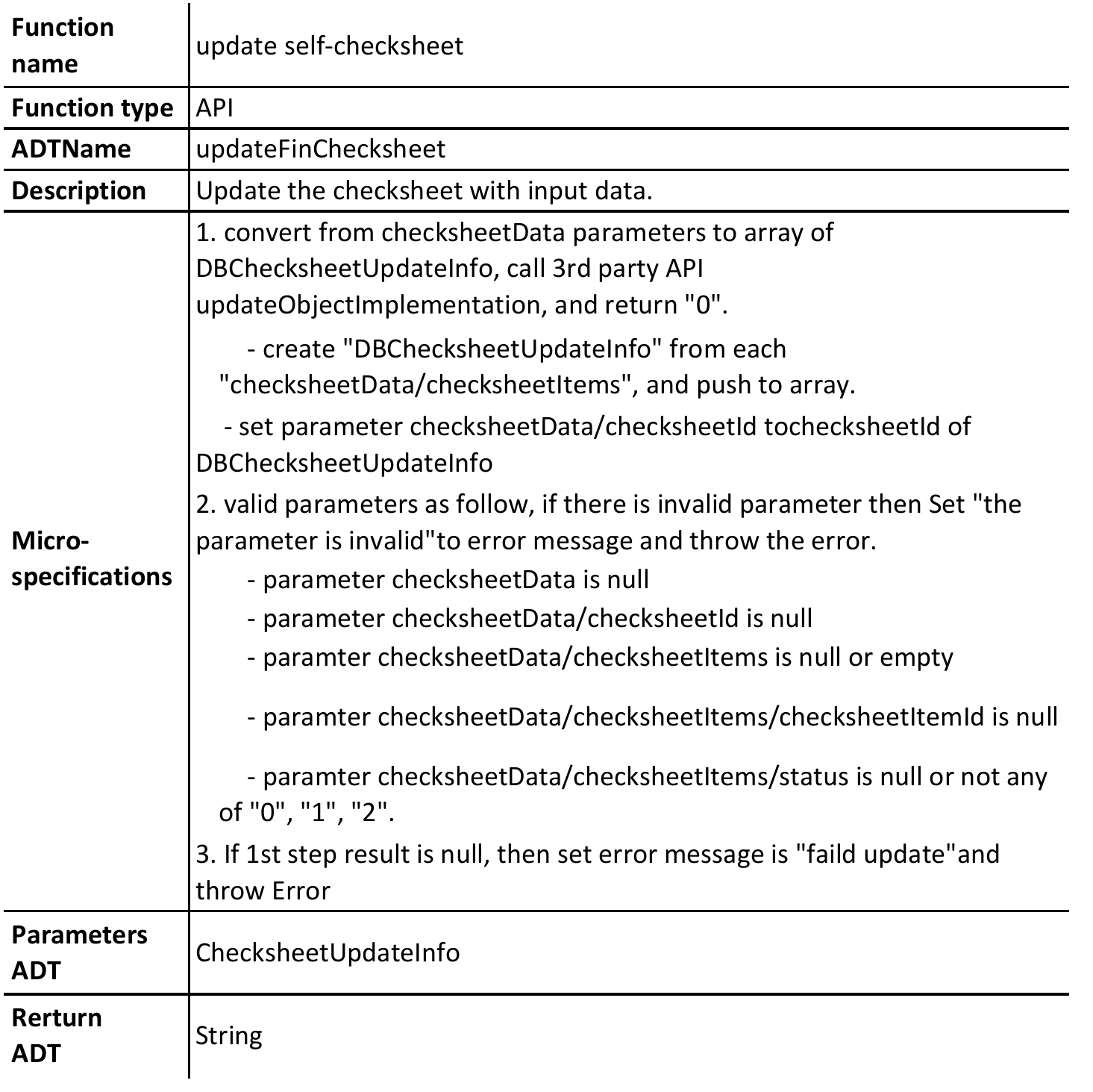}

\caption{The function~\texttt{updateFinChecksheet} API was described through three micro-specifications created by the dedicated software engineer. Each micro-specifications was then mapped to a microtask. A crowd worker completed each microtask, and a second crowd worker reviewed the work.} 
\label{fig:taskExample}
\end{figure}

Next, in the detailed design step, the Functions API and UI components specifications were decomposed by the dedicated software engineer into a set of micro-specifications.
For example, in Figure~\ref{fig:taskExample}, the Function API specification was decomposed into three micro-specifications. The dedicated software engineer generated three types of artifacts: abstract data types, Functions API (backend), and UI components micro-specifications(frontend).

\begin{figure}
\centering
\includegraphics[width=\columnwidth,keepaspectratio, clip]{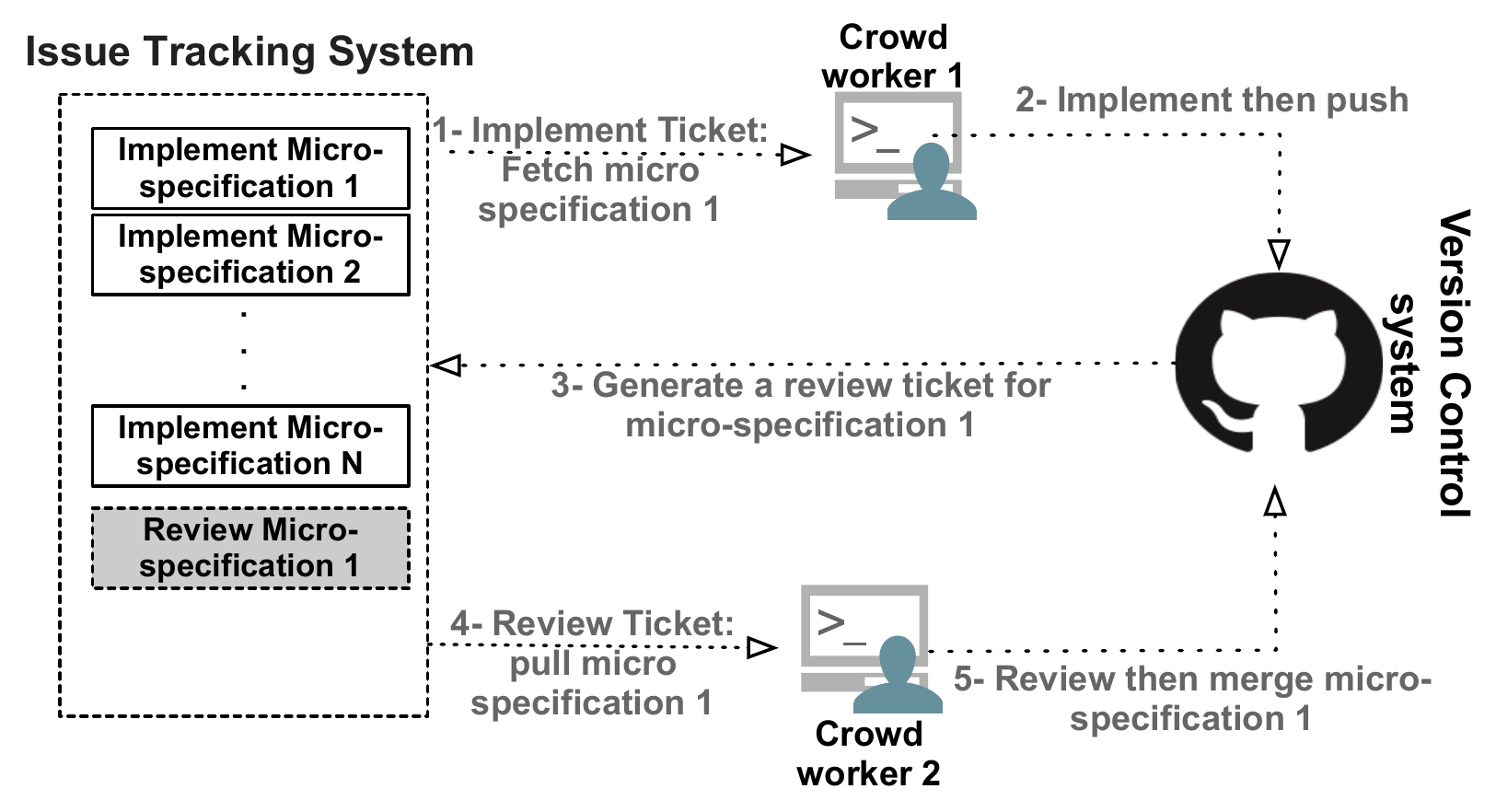}

\caption{An example of work being completed by two crowd workers within the microtask programming workflow. The dedicated software engineer first decomposed each specification into multiple micro-specifications, creating a  ticket describing each micro-specification to implement. Crowd worker one may then fetch a ticket from the issue tracking system (ITS), complete the work, push the source code, and create a pull request in the version control system. Crowd worker two may continue the work, fetching the pull request and reviewing the source code implemented by crowd worker one. If crowd worker two approves the work, they may then merge the pull request into the version control system (VCS).}
  \label{fig:exampleOfTwoCrowd}
\end{figure}

In the implementation and unit testing step, each of the micro-specifications were then implemented by the crowd workers. Crowd workers followed Test Driven Development (TDD) to implement micro-specifications. For example, Figure~\ref{fig:exampleOfTwoCrowd} depicts an example of a micro-specification being fetched, implemented, and pushed by one crowd worker and reviewed by a second crowd worker. \par

In the testing step, the dedicated software engineer created a test suite. They then tried to build all of the components together. There were several defects that were found. In some cases, defects were fixed by the dedicated software engineer, while in others they were fixed by the crowd workers.\par

\begin{figure}[h]
\centering
\includegraphics[width=\columnwidth,keepaspectratio, clip]{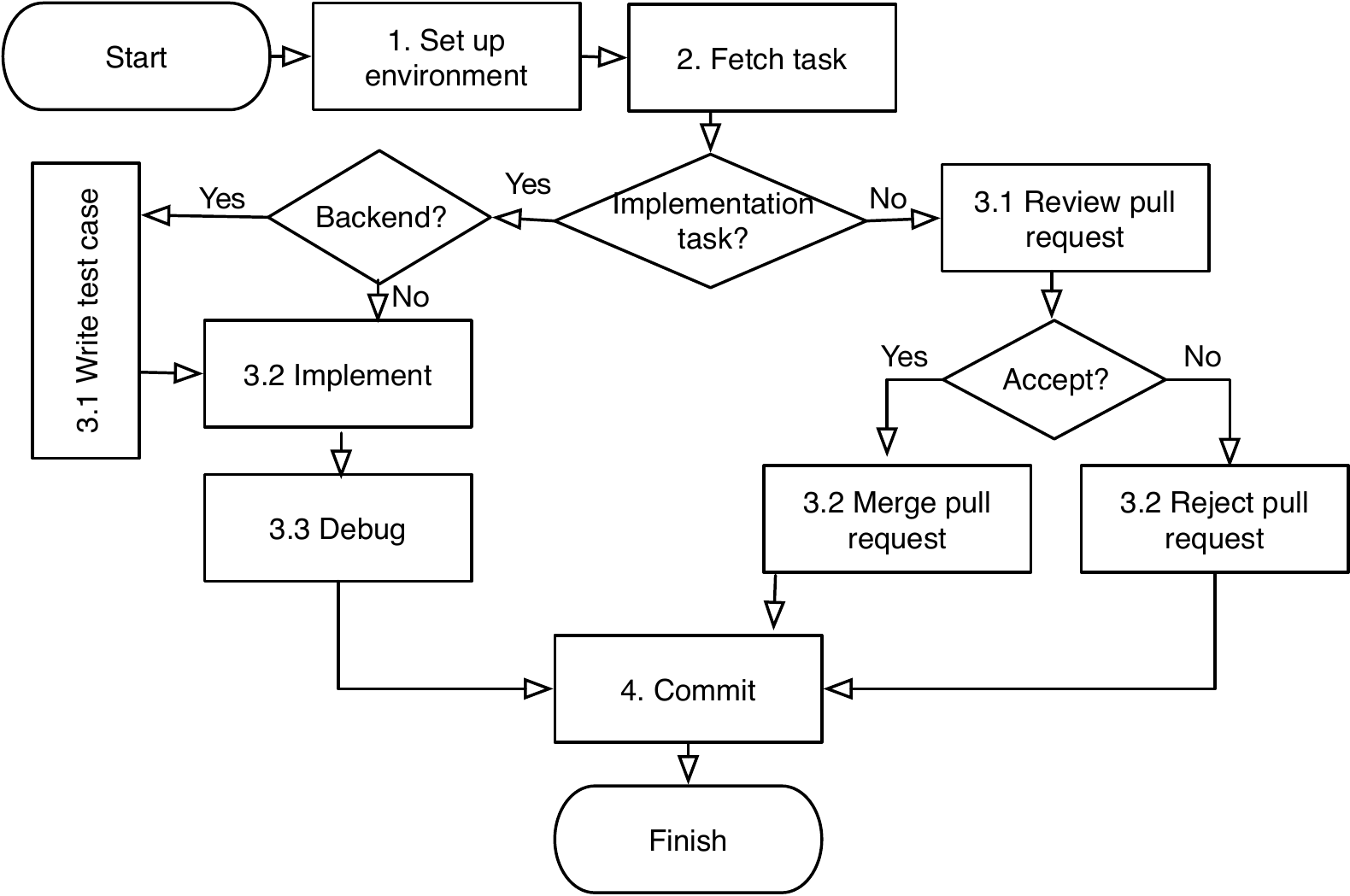}

\caption{In the microtask programming workflow, each crowd worker worked through a series of steps. Workers first fetched a ticket, which included backend and frontend tickets as well as review tasks. For implementation tasks, workers then completed several programming steps before committing their work and finishing the microtask.}
  \label{fig:workflowInFlowchart}
\end{figure}

\subsection{Workflow and Tools for Crowd Workers}

Figure~\ref{fig:workflowInFlowchart} describes microtask workflow used by the transient crowd workers in contributing to the project.  
Figure.~\ref{fig:exampleOfTwoCrowd} depicts an example of how micro-specifications were fetched, implemented, and committed. Two key tools were used: an Issue Tracking System (ITS) and  Version Control System (VCS). Participants used the ITS to manage the status of the work and the version control system to manage changes to code.\par

In the microtask programming workflow, the software engineer first decomposed the specifications generated by the requirements engineer into micro-specifications and issued a ticket for each micro-specification in the ITS (an example of a micro-specification is shown in Figure.\ref{fig:taskExample}). Next, a crowd worker fetched a ticket from the ITS. Crowd workers selected the work that they chose to do. They then implemented the micro-specification described in it by using the template imported from the VCS.  After completing the implementation, the crowd worker then pushed the source code and created a pull request. A second crowd worker then reviewed the contribution. After fetching the ticket in the form of a pull request, the second worker reviewed the source code. If they approved the work, they then merged the pull request.\par

Crowd workers also made use of the ITS to communicate project knowledge. Crowd workers asked the dedicated software engineer questions about the micro-specifications posted on the ITS. If the designer found the question to be broadly applicable, they shared it with all of the crowd workers by posting it to an internal wiki. When crowd workers began contributing to the project, they were then asked to first read the internal wiki. \par

\section{Results}
In the following section, we first report on the outcome of the project, describing the activities that occurred and the results of the project. We then report data  on the productivity of crowd workers within the project. Finally, we examine the project participant's perceptions of the suitability and effectiveness of applying microtask programming in a company setting through collected interview and survey data. 

\subsection{Project Activities}
Work in the project occurred in five phases over a seven week period: basic design (one week), detailed design (one week), sprint 1 (three weeks implementation and unit testing, and three-day system testing), and sprint 2 (one week implementation and unit testing, and two days system testing). Transient crowd workers contributed to the project in both sprints. The dedicated software engineer was the same in the two sprints. 
In each sprint, crowd workers were assigned to either frontend or backend work. \par

To onboard onto the project, crowd workers attended a kickoff meeting, read content on a dedicated wiki, and set up their computer environment. Before beginning implementation work, all of the crowd workers attended a 30-minute kickoff meeting in which the requirements engineer and the dedicated software engineer offered an overview of the microtask programming workflow. Rather than impart knowledge about the project itself, the briefing focused exclusively on the use microtask programming. After the kickoff meeting, when the crowd workers were ready to begin working on the tasks, they setup their environment. To do so, they read the information in the wiki which we had prepared beforehand. The wiki instructed them about how to set up the environment and how to use it for work. Rather than use a pre-defined environment within a virtual machine, crowd workers instead set up their own laptop or desktop computers from scratch. Crowd workers reported that it took approximately two hours to set up their environment on average, including reading and understanding the contents of the wiki and configuring their computer’s environment. The environment consisted of the Visual Studio IDE, web browser, Git VCS, node.js and some JavaScript libraries. Crowd workers spent approximately 30 to 60 minutes configuring their computer environment. \par

After the crowd workers completed their work in the implementation step in sprints 1 and 2, the dedicated engineer tested the project. 
The contributions from each microtask were composed into an assembled program and tested through a test suite to evaluate if the system satisfied its requirements. System testing was conducted using standard company practices, where test cases were implemented based on the requirements described in the use case and screen transition artifacts. 
Two types of defects were found. Some defects were related to the micro-specifications created by the dedicated software engineer. Other defects were related to the implementation of the micro-specifications by the crowd. The dedicated software engineer fixed all of the defects.


All of the defects uncovered were minor issues. The dedicated software engineer 
confirmed that the specifications were completely implemented. Based on the successful test results, the project entered the next phase of user acceptance testing.

\par

\begin{table}[h]
\caption{The number of components, micro-specifications, active crowd workers, and dedicated software engineers for each sprint. Some crowd workers contributed to both the backend and frontend. }
\label{tab:workDetail}
\resizebox{\columnwidth}{!}{%
\begin{tabular}{@{}c c c c c c @{}}
\textbf{} & \textbf{} & 
& \textbf{\begin{tabular}[c]{@{}c@{}} \#micro-\\specification\end{tabular}} & \textbf{\begin{tabular}[c]{@{}c@{}}\#Crowd\\ Workers\end{tabular}} & \textbf{\begin{tabular}[c]{@{}c@{}}\#Dedicated\\ SE\end{tabular}} \\ \midrule

\multicolumn{1}{l}{\multirow{2}{*}{\textbf{Sprint 1}}} & \textbf{Backend}   & 5 Functions API  & 20  & 5 & \multirow{2}{*}{1}  \\ \cmidrule(lr){2-5}

\multicolumn{1}{l}{}& \textbf{Frontend} & 3 UI Components& 7& 3&      \\ \midrule
\multicolumn{1}{c}{\multirow{2}{*}{\textbf{Sprint 2}}} & \textbf{Backend}  & 4 Functions API                         & 18& 3& \multirow{2}{*}{1}\\ \cmidrule(lr){2-5}
\multicolumn{1}{c}{}& \textbf{Frontend} & 2 UI Components& 5& 2&\\ \bottomrule

\multicolumn{2}{c}{\textbf{Total of both sprints}}& \multicolumn{1}{c}{9 Functions API, 5 UI Components}& \multicolumn{1}{c}{50}& \multicolumn{1}{c}{6}& \multicolumn{1}{c}{1}\\
\end{tabular}%
 }
\end{table}

\begin{table}
\caption{The lines of code written by the crowd workers and the dedicated software engineer by sprint. }
\label{tab:LOC}
\resizebox{\columnwidth}{!}{%
\begin{tabular}{@{}cccccc@{}}

\textbf{}&
\multicolumn{1}{c}{\textbf{}}  &
\multicolumn{1}{c}{\textbf{\begin{tabular}[c]{@{}c@{}} \#LOC by \\Crowd Workers\end{tabular}}} & \multicolumn{1}{c}{\textbf{\begin{tabular}[c]{@{}c@{}} \#LOC by \\Dedicated SE\end{tabular}}} & \multicolumn{1}{c}{\textbf{\begin{tabular}[c]{@{}c@{}} \#Total \\of LOC\end{tabular}}} \\
\cmidrule[\heavyrulewidth]{1-5}
\multicolumn{1}{c}{\multirow{4}{*}{ \rotatebox[origin=c]{90}{\textbf{Sprint 1}}}} &

\multicolumn{1}{l}{ \begin{tabular}[c]{@{}c@{}}Backend \\Implementation  and Testing\end{tabular}}  &\multicolumn{1}{c}{905} & \multicolumn{1}{c}{245}& \multicolumn{1}{c}{1150} \\ \cmidrule(l){2-5} 

\multicolumn{1}{c}{} & \multicolumn{1}{l}{\begin{tabular}[c]{@{}c@{}}Frontend \\Implementation  and Testing\end{tabular}}&  \multicolumn{1}{c}{686}&\multicolumn{1}{c}{1569}& \multicolumn{1}{c}{2255}\\ 
\cmidrule(l){2-5}

\multicolumn{1}{c}{}& \multicolumn{1}{l}{System Testing} & \multicolumn{1}{c}{-}&\multicolumn{1}{c}{468}& \multicolumn{1}{c}{468}
\\ \cmidrule(l){2-5} 

\multicolumn{1}{c}{}& \multicolumn{1}{l}{\textbf{Total}}& \multicolumn{1}{c}{1591}& \multicolumn{1}{c}{2282}& \multicolumn{1}{c}{3873}\\ 
\cmidrule[\heavyrulewidth]{1-5}



\multicolumn{1}{c}{\multirow{4}{*}{ \rotatebox[origin=c]{90}{\textbf{Sprint 2}}}} &

\multicolumn{1}{l}{\begin{tabular}[c]{@{}c@{}}Backend \\Implementation  and Testing\end{tabular}}& \multicolumn{1}{c}{800}& \multicolumn{1}{c}{245}& \multicolumn{1}{c}{1045}\\ \cmidrule(l){2-5}

\multicolumn{1}{c}{}& \multicolumn{1}{l}{\begin{tabular}[c]{@{}c@{}}Frontend \\Implementation  and Testing\end{tabular}}& \multicolumn{1}{c}{500}& \multicolumn{1}{c}{1859}&\multicolumn{1}{c}{2359}\\ 
\cmidrule(l){2-5} 

\multicolumn{1}{c}{}& \multicolumn{1}{l}{System Testing} & \multicolumn{1}{c}{-}& \multicolumn{1}{c}{671}&\multicolumn{1}{c}{671}\\ 
\cmidrule(l){2-5}

\multicolumn{1}{c}{}& \multicolumn{1}{l}{\textbf{Total}}& \multicolumn{1}{c}{1300}& \multicolumn{1}{c}{2775}& \multicolumn{1}{c}{4075}\\ 
\cmidrule[\heavyrulewidth]{1-5}

\multicolumn{2}{l}{\textbf{Total of both sprints}}& \multicolumn{1}{c}{2891}& \multicolumn{1}{c}{5057}& \multicolumn{1}{c}{7948}\\ 


\end{tabular}%
}

\end{table}

\subsection{Crowd Worker Productivity}
Across the two sprints, the crowd workers implemented 9 Functions API and 5 UI Components through 50 microtasks. The work completed is listed in Table.~\ref{tab:workDetail}. In sprint 1, five crowd workers worked on the backend to develop five functions. On average, each crowd worker implemented 4 microtasks in the first sprint. Three crowd workers worked on the frontend to develop three UI Components based on seven microtasks created. Two crowd workers worked on both the frontend and backend. In sprint 2, three crowd workers worked on the backend to develop four functions. Each crowd worker implemented an average of 6 microtasks. Two crowd workers worked on the frontend to develop two UI Components based on five microtasks. In contrast to sprint 1, each crowd worker worked exclusively on either the frontend or the backend. While seven crowd workers were initially invited, one crowd worker was not able to contribute to the project as they did not have any slack time. \par

To assess the output of the crowd and dedicated software engineer, we measured the lines of code (LOC) produced in each sprint.  Table~\ref{tab:LOC} summarizes the results. In sprint 1, the crowd workers and the dedicated software engineer developed 1,591 LOC and 2,282 LOC, respectively. In sprint 2, the crowd workers developed 1,300 LOC and the dedicated software engineer developed 2,775 LOC. The dedicated software engineer implemented more LOC in this sprint as they implemented templates and packages in the frontend. 
The final project consisted of approximately 8,000 LOC, of which about 2,900 LOC (36\%) were implemented by the crowd workers. The average implementation microtask involved 58 lines of code 
.\par

We also used the timestamp data recorded in the tickets of the issue tracking system to examine the productivity of the crowd workers. Table.~\ref{tab:contributionsDetails} lists the average working time and the number of fetched tickets, and micro-specification for each crowd worker. Frontend tasks took more time than backend tasks. In total, 50 microtasks were generated and 66 were submitted by crowd workers. Microtasks may be submitted more than once when a pull request was rejected. In total, 50 microtasks were generated, and crowd workers submitted 66 microtasks. On average, 75\% of implementation microtasks were accepted.  \par 

There are two reasons why average times were in some cases more than one hour. First, some tasks were too big or too complex to do in an hour. %
The second is that crowd workers sometimes fetched a ticket and switched to another task. Ticket timestamps sometimes indicated that the crowd worker fetched tasks in the evening and did not start working on it until the following day. On several occasions, one crowd worker fetched a task and only began it later. As a result, Crowd Worker Two in the backend had an average work time of 11hr, 35 min. While workers had the ability to pause the time in the ITS, some forgot to do so.\par

\begin{table}
\centering
\caption{The number and average completion time of microtasks by each crowd worker (CW)}
\label{tab:contributionsDetails}
\resizebox{\columnwidth}{!}{%
\begin{tabular}{ccccccccc}
\textbf{}  &
\textbf{}  &
\multicolumn{3}{c}{\textbf{Implementation Microtasks} } &
\multicolumn{2}{c}{\textbf{Review Microtasks} } \\ 
\cline{3-7}

\textbf{}& 
\begin{tabular}[l]{@{}l@{}}\rotatebox[origin=l]{90}{\textbf{Worker}} \end{tabular} &
\begin{tabular}[c]{@{}c@{}}\textbf{Average Time}\\\textbf{(hh:mm)} \end{tabular} &
\begin{tabular}[c]{@{}c@{}}\textbf{\#Implemented}\end{tabular} & \begin{tabular}[c]{@{}c@{}}\textbf{\#Accepted}\end{tabular} & 

\begin{tabular}[c]{@{}|c@{}}\textbf{Average Time}\\\textbf{(hh:mm)} \end{tabular} & 
\begin{tabular}[c]{@{}c@{}}\textbf{\#Reviewed}\end{tabular}\\ 
\cmidrule{1-7}

\multirow{6}{*}{
\rotatebox[origin=c]{90}{\textbf{Backend}}
} 
 & \textbf{CW 1}  & 00:45 &  11 & 8 & 00:13 & 3 \\\cmidrule{2-7}
 & \textbf{CW 2}  & 11:35 &  12 & 11 & 00:43  & 7 \\\cmidrule{2-7}
 & \textbf{CW 3}  & 00:36 &  2 & 2 & 00:06 &  8 \\\cmidrule{2-7}
 & \textbf{CW 4}  & 00:18 &  8 & 6 & 00:08 &  13 \\\cmidrule{2-7}
 & \textbf{CW 5}  & 03:23 &  7 & 4 & -- & --  \\ \cmidrule{2-7}
 & \textbf{CW 6}  & 02:17 &  7 &7 & 00:21 &13 \\ \cmidrule{2-7}
 \cmidrule[\heavyrulewidth]{1-7}

\multirow{5}{*}{
\rotatebox[origin=c]{90}{\textbf{Frontend}}
} 
 & \textbf{CW 1}  & 01:46 & 6 & 2 & 00:13  & 1 \\\cmidrule{2-7}
 & \textbf{CW 2}  & 05:29 & 6 & 4 & 00:46  & 3 \\\cmidrule{2-7}
 & \textbf{CW 3}  & 00:38 & 2 & 2 & 00:08  & 7 \\\cmidrule{2-7}
 & \textbf{CW 5}  & 08:52 & 2 & 2 & 00:06  & 1 \\ \cmidrule{2-7}
 & \textbf{CW 6}  & 09:03 & 3 & 2 & 00:48  & 7 \\\cmidrule{2-7}
\cmidrule[\heavyrulewidth]{1-7}
\textbf{Total}& --  & -- &  66 & 50 & -- &  63 
\end{tabular}
}
\end{table}

\begin{figure*}
\centering
\includegraphics[width=\textwidth,keepaspectratio, clip]{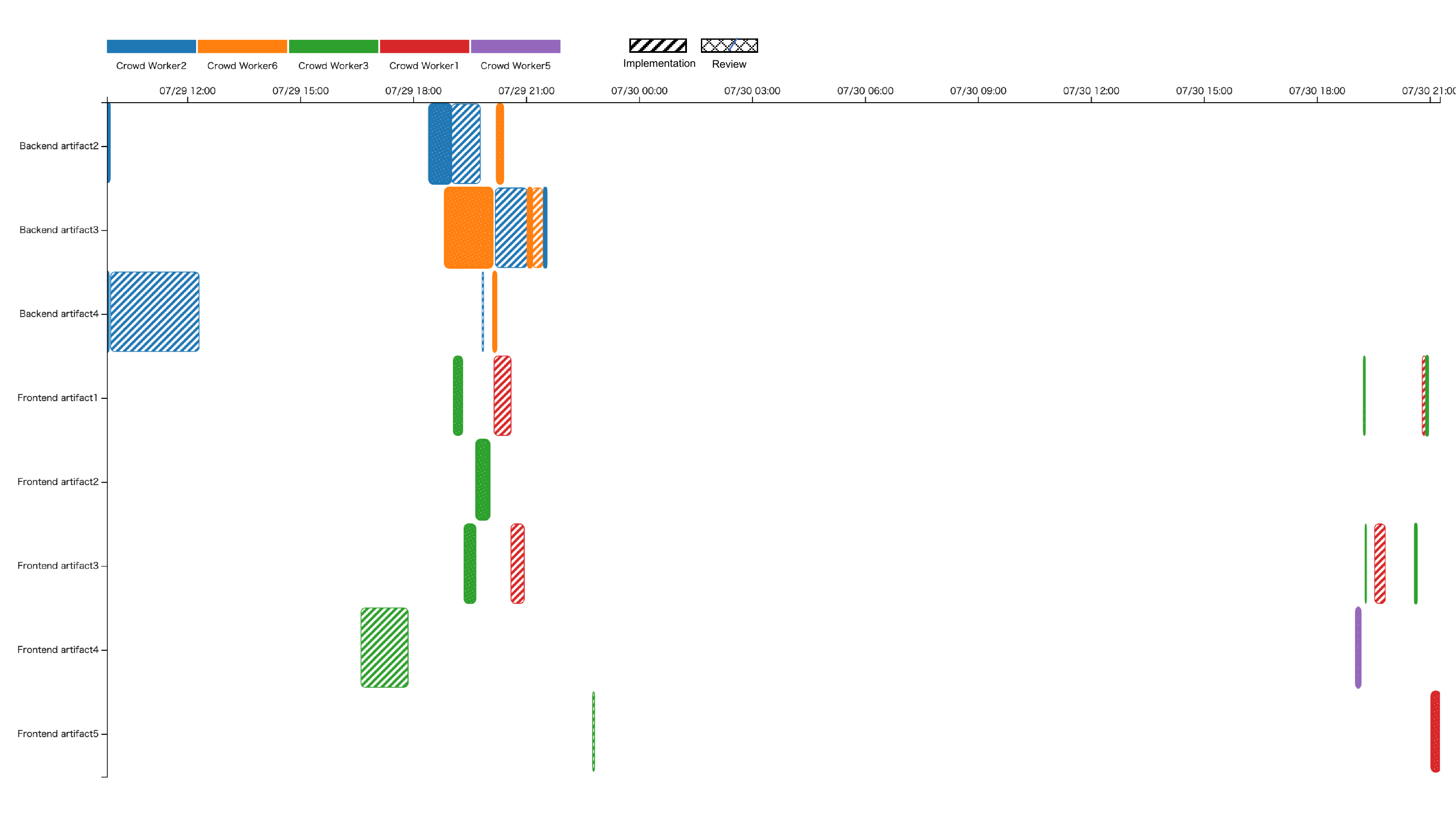}
\caption{A visualization of crowd workers' switching between artifacts during the project. The vertical axis is the artifact workers were contributing to and the horizontal axis time. Each bar in the plot corresponds to a single microtask (either implementation or review microtask). Crowd workers are indicated through colors. The bars' pattern indicates the task type, with striped bars indicating implementation tasks and filled bars indicating review tasks. }
  \label{fig:switches}
\end{figure*}

Figure~\ref{fig:switches} depicts the work done by each crowd worker, from the time they fetched a microtask until they submitted it.
For example, on Backend artifact 3, Crowd Worker 6 first completed a Review microtask in about one hour, Crowd Worker 2 completed an implementation microtask in about one hour, Crowd Worker six then completed Review and Implementation microtasks in succession, and finally Crowd Worker 2 completed a a Review microtask.

Crowd workers were prohibited from fetching multiple microtask at the same time, although they sometimes forgot to close the ticket and started a new task.  Work on an individual artifact often switched between different crowd workers. Work often clustered around points of time with more intense activity. Most crowd workers contributed during the daytime or working for 2 or 3 hours in the evening. \par

We did not precisely record the hours worked on the project by the dedicated requirements engineer and software engineer (Fig. ~\ref{fig:GeneralOverview}). However, we were able to estimate the time. On average, the dedicated engineers worked for 8 hours per day. The requirements engineer spent one week on the project (40 hours).  The software engineer spent 40 hours in the detailed design phase. In those 40 hours, the software engineer exclusively worked on converting basic design diagrams to function specifications and preparing microtasks. In the last phase, the software engineer spent an additional 40 hours conducting system testing.\par

\subsection{Perceptions of Use by Crowd Workers}
To understand the perceptions of the crowd workers about the suitability and value of microtask programming in a company setting, we gathered data through two methods. One week after the completion of the project, we asked the 6 crowd workers to complete a short questionnaire about their experiences with microtask programming inside a company. The questions focused on onboarding challenges, the granularity of microtasks, the freedom to choose a task, the motivation of working in the microtask programming approach, and communication among workers. 
We then later conducted 15-30 minute semi-structured interviews with each of the six crowd workers.  The open-ended questions focused on onboarding challenges, the granularity of microtasks, the freedom to choose a task, the motivation of working in the microtask programming approach, and communication among workers.

\subsubsection{Questionnaire Results}

\begin{figure}[h]
\centering
\includegraphics[width=\columnwidth,keepaspectratio, clip]{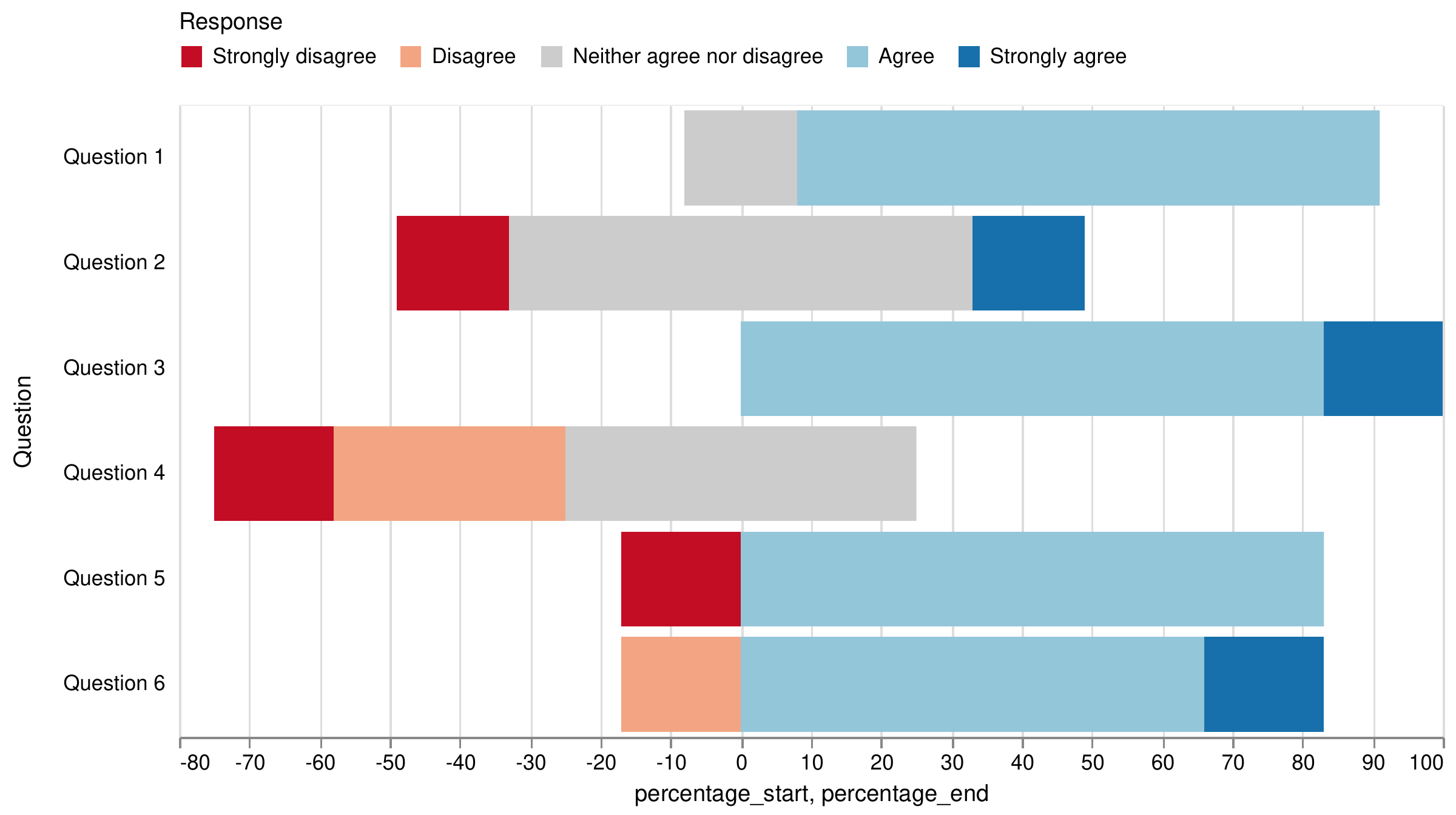}
\caption{Perceptions of microtask programming by crowd workers.\\
 \textmd{Question 1: I think the microtask programming style reduces my onboarding cost in comparison to the generic model (e.g., agile model).\\
Question 2: I think that completing a single task on an individual small artifact is easily performed.\\
Question 3: I find that it is comfortable to select (or skip) my task by myself.\\
Question 4: I think that it is easy for me to stay motivated in the microtask programming style.
\\
Question 5: I think it is inconvenient not to have face to face communication with the software designers.
\\
Question 6: I think it is inconvenient not to have face to face communication with other crowd workers.
}
}
  \label{fig:likertResult}
\end{figure}

The results from the questionnaire are listed in Figure~\ref{fig:likertResult}.
The crowd workers mostly agreed that microtask programming reduced the time required to onboard onto a project. Each crowd worker took on average only 150 minutes (30 minutes for the kickoff meeting and 120 configuring the environment) to join the project. As much of this time was dedicated to understanding the microtask workflow and configuring the related tools, this time would likely be less were crowd workers to participate in another microtask programming project. \par

Workers reported varied opinions on working with microtasks. Most agreed with ability to select or skip each microtask. However, crowd workers reported that staying motivated was often hard. Crowd workers did not find the lack of face to face communication to be challenging.\par

\subsubsection{Interview Results}

Most of the crowd workers felt that microtask programming decreased onboarding costs. One reported that the wiki which had been prepared beforehand effectively reduced the onboarding cost by including
detailed information on how to set up the environments and how to use it for work.  Other workers reported difficulties onboarding. One reported that it takes time to learn rules and appropriate methods of work. Another worker reported that a pre-defined environment such as a virtual machine would have considerably reduced onboarding costs.\par

Most workers had neutral opinions about the clarity of the microtasks. For three workers, some micro-specifications were not clear. For example, one reported that: "Text-based specification is hard to understand. If the specification had included figures and/or models, it would have been better." Workers reported following two approaches for resolving unclear micro-specifications. The first one was skipping the ticket in the ITS and beginning work on another ticket. The second approach was to create a ticket in the ITS to ask the dedicated software engineer a question. \par

Crowd workers were comfortable selecting microtasks themselves. Two workers reported that micro-specifications had many details about the logic that were helpful. Two workers reported that implementation microtasks were a good fit for the backend, as they were neither too small nor too big. Workers also reported concerns about implementation microtasks. \par

One worker reported concern that the sequence of microtasks was not specified, as how a micro-specification was implemented might impact the implementation of another micro-specification. 
Another worker reported being concerned that a behavior implemented earlier might no longer work correctly after another micro-specification was implemented.
When a crowd worker implemented a micro-specification and wrote a test for it, they focused only on the micro-specification they were assigned. When they wrote the test, it might pass. But it might later fail after another crowd worker implemented other functionality. 
For example, Crowd Worker 1 wrote a test for micro-specification A where he invoked a function with empty parameters. At the time, the test passed. However, according to micro-specification B, the function must be invoked with parameters. When Crowd Worker 2 implemented this behavior, the test for micro-specification A then failed. In this way, unit tests helped to facilitate coordination.
\par

Crowd workers had different views about the value of review microtasks. Most thought review microtasks were easy to understand, as the scope was narrow. However, when the micro-specification was incorrect or unclear, this created challenges. To resolve this, crowd workers had to write on the ticket to communicate about the specification. They instead wished to be able to communicate directly with the dedicated software engineer who had authored the micro-specification.
Moreover, one worker reported that, while he also does review tasks in regular software development, review microtasks are different. He reported that: "Usually, the viewpoints of [the] review are based on syntax such as coding rules, format, and so on. But in this project, I had to check whether the code meets behavior; it takes a longer time than usual." Another crowd worker complained that his work was rejected several times because the code violated code rules.\par

All crowd workers reported that understanding a micro-specification and implementing backend microtasks could be completed in less than one hour, with a wide range of responses ranging from 20 to 60 minutes. They reported that completing frontend microtasks might take more than one hour, as the microtasks were larger. \par

Crowd workers were transient, and they could work on the project in their slack time. One of the workers reported a preference to complete multiple microtasks at a time. He told us he prefers to complete 5 or 6 microtasks in one session for a couple of hours. \par

Based on the results of the questionnaire, we investigated the challenges for crowd workers in staying motivated in microtask programming. Crowd workers reported several reasons. Two reported that they did not feel that they belonged to the project. These crowd workers reported that if a microtask had been assigned to them and the microtask had a due date, they would have been more motivated. Two other workers reported that it was not clear if there remained incomplete microtasks. If someone had asked them to complete the incomplete microtasks which, they would have done that. Another reported that since he thought someone else would complete microtasks, he did not feel that he should do that. \par

Crowd workers reported several suggestions for increasing the project velocity and the motivation of crowd workers. Two workers suggested that instead of arbitrarily choosing microtasks they should instead be assigned. Workers would then know that they had a microtask that they must complete before its due date. Three crowd workers shared that if there was an incentive like a score or money, they would work harder. Another crowd worker suggested dedicating a specific day to microtask programming. \par

The interview results confirm many of the questionnaire results. Most crowd workers did not have problems with the lack of face-to-face meetings with crowd workers. Four workers reported that they only needed communication in the onboarding phase of the project or their first microtasks. After that, they did not have specific questions. Thus they felt they did not need any specific tools for communication. They believed that workflow and tasks were clear enough. On the other hand, two participants reported it would be better if all communication were in a shared place like a question and answer tool. They felt that sharing information might increase their productivity. One participant reported: "... I prefer chat to [a] ticket. I feel that it is a good point to record the history of the actor's actions using tickets. However, in many scenes, I need quick answers from software designers. So, the chat is better than a ticket." \par

\subsection{Perceptions of Use by Dedicated Software Engineer}
We conducted an interview with the dedicated software engineer to gather feedback about the challenges he faced when he decomposed tasks into microtasks and when he assembled the contributions into a finished application. \par

Although the software engineer could complete all of the tasks by himself, the software engineer reported that his time commitment was reduced. While crowd workers were working on the implementation microtasks, the software engineer used that time to prepare the integration test and micro-specification for the next sprint. As he had designed the project, he was able to complete the tasks in a shorter time than the crowd workers. It is not clear if someone else had created the tasks and the software engineer was not involved in the design phase how much it time it would have taken for the dedicated software engineer to complete all of the tasks.\par

Manually decomposing tasks into microtasks  was challenging. In the backend, he tried to decompose a specification of a module into a set of testable, smallest behaviors. 
It was challenging to describe complex behaviors using only text. In the frontend, he tried to decompose the web pages into a set of components. Each was the smallest meaningful unit of the UI element. It was challenging to consider what the minimum information needed to implement a microtask was. The software engineer struggled to decide the level of detail necessary for the micro-specifications. It was difficult for the software engineer to predict how much time the crowd workers would spend understanding the micro-specifications he wrote. He also thought that the reason why the tasks took much more time than he expected was the size of the microtasks. He described much of the contents in the UI component’s specifications. So these required much time to implement.\par

Aggregating microtasks and conducting integration tests was not straightforward for the software engineer. There were not automated continuous integration tools, which added much time.  He suggested that the test results might be used to monitor project progress. 


\section{Limitations and threats to validity}

When we design any case study, care should be taken to mitigate threats to validity~\cite{yin2003a}.\par

Construct validity addresses the degree to which a case study aligns with the theoretical concepts used. To reinforce construct validity, there are three ways: using multiple sources of reliable evidence, establishing a chain of evidence, and having key informants review reports of a draft case study~\cite{yin2003a}. In this case study, we used only one source from one web application system development project. To establish a chain of evidence, ITS and VCS were used to maintain a record of all data of the study. Finally, outside researchers studying crowdsourcing software engineering were involved in the study and reporting of study results.
In this case study, we make no causal inferences, so internal validity is not a concern.\par
External validity is the ability of a case study’s findings to generalize to the broader population of interest~\cite{yin2003a}. A possible threat to external validity is that we only analyzed one project. Microtask programming is not inherently domain specific and would be relevant for other projects beyond developing web applications. Both of the key supporting tools (ITS and VCS) in the case study are widely used in software development projects. This supports the external validity of our study.\par

Reliability is the ability to repeat a study and observe similar results~\cite{yin2003a}. To reinforce our study’s reliability, we defined and documented the microtask programming workflow and the approach for assigning microtasks to crowd workers. By using the same process and tools, other researchers or practitioners may replicate the case study in their own context.\par

\section{Discussion}
In this paper, we investigated the potential of applying microtask programming in a company setting, reporting on a project undertaken at NTT to build a web application through microtask programming. 
Japanese IT vendors such as NTT face a serious IT talent shortage. Microtask programming may help to address this by increasing the fluidity of developer assignments within large organizations. \par

In adopting microtask programming,  there were four key ways in which development work differed from standard practice: (1) a finer granularity of work, (2) less face to face communication, (3) associating contributions with artifacts, and (4) enabling developers primarily assigned to a different project to contribute in their free time. Contributions were smaller than typical, focusing on implementing or reviewing the behavior specified in an individual micro-specification. Rather than participate in the typical daily face to face or conference meetings to coordinate, crowd workers instead collaborated by reading tickets describing their work. Rather than work on tickets encompassing an issue spanning multiple source code artifacts, crowd workers instead worked on an individual artifact, reducing necessary project knowledge. And rather than work exclusively on a dedicated software project, seven crowd workers made contributions to a second project in their available free time. These differences led to a different style of work.\par

There were also differences in the nature of the review tasks. The focus of code reviews is traditionally on conventions, such as coding rules and format. In microtask programming, workers were also responsible to check whether the code satisfied its expected behavior. As a result, review tasks took longer than usual. 

Adapting microtask programming did not require a change to the project life-cycle . Microtask programming was organized in the same four key project steps typically used by the organization (Fig. ~\ref{fig:GeneralOverview}). The key difference was that microtasking added a new level of detail in the detailed design step, where the specifications were decomposed into micro-specifications. \par

The results demonstrate the potential for microtask programming to increase the fluidity of project assignments within an organization. The project met its objectives and successfully completed its system testing. Crowd workers assigned to other projects were able to contribute in their slack time and largely felt that the onboarding costs were reduced.\par

Prior to beginning the project, there were concerns about the quality of the code and the project's time to market. It was not clear how much the output of the crowd could be trusted. The project was built and tested to evaluate if the system satisfied its requirements. The system testing revealed only a few defects, which was considered quite successful.
There were also concerns about the efficiency of microtask programming. It was not clear how much time would be needed to complete a project through microtask programming. However, in less than 2 months, the project completed 9 functions and 5 UI Components. This alleviated concerns that microtask programming might be excessively slow.  \par

The project also revealed several challenges with applying microtask programming. The crowd workers reported that they found staying motivated to be harder. 
Crowd workers found the style of work used in microtasking to be unfamiliar and different. Motivation may also have been reduced because microtask programming was not typical practice and workers did not feel they needed to be as productive. Workers did not directly gain anything for their contributions, further reducing their motivation. \par

From the viewpoint of project management, it was difficult to monitor the status of the crowd workers because they were assigned to other projects. Only when they were not busy in their own project could they make contributions to the project. This made it hard to anticipate when and how often they would be able to contribute. As a result, estimating progress and handling risk in the project were very difficult.





 

While project participants perceived onboarding overhead to be reduced, there was still considerable overhead involved for crowd workers to get started and begin contributing. 
At the beginning of the project, workers were confused about the new concept and workflow. Workers also did not have their regular face to face meetings. It may be that after workers become more comfortable and familiar with microtask programming, some of this overhead may decrease. Moreover, while microtask programming often includes a pre-configured development environment, workers did not benefit from this in our study. Offering this might  reduce onboarding overhead.\par

In this project, crowd workers manually selected microtasks from an issue tracking system. 
We are planning to develop techniques to generate these assignments, by considering crowd worker's main area of work and expertise and their current daily tasks.\par



In the project, the dedicated software engineer's workload increased during the detailed design step in which they created the micro-specifications. As there was no systematic method available for completing this work, the dedicated software engineer had to manually determine how to do this through trial and error, which increased their workload. 
This effort might be reduced through tool support, creating more detailed guidelines, or by adapting the workflow to find new ways to use the crowd. \par

Microtasking was only adopted for parts of the project which required less project knowledge to complete or that were not complex. Tasks that required more knowledge still involved a dedicated software engineer, as they required a level of knowledge that crowd workers did not have. This included 1) designing the system architecture, 2) designing and configuring the database, 3) completing tasks that involved screen transitions, 4) completing tasks which required infrastructure knowledge. 
In addition, the project's scope was limited to a development phase, rather than encompassing other phases such as maintenance. 
Better understanding how to employ microtasking in more complex programming tasks, when focusing on non-functional or crosscutting considerations such as performance, and in phases such as maintenance are essential questions for future work. 
\par



\section{Conclusion}
In this paper, we reported an industrial case study of the application of microtask programming to a web application development project. Work to implement a system was decomposed into a set of micro-specifications, which were then implemented and reviewed by the crowd. Crowd workers were primarily assigned to other projects and worked in their free time. A system with approximately 8,000 LOC in total was built, of which approximately 35\% was implemented and reviewed by six crowd workers. Individual contributions made by crowd workers were small, averaging about 58 LOC. \par

The project results suggest the promise of microtask programming for making the assignment of developers to projects more fluid.  Based on the results from this project, high-level management at NTT has realized the potential benefits. We plan to apply the approach to not only other development projects but also to system operation and maintenance projects. We are also planning to develop a tool for automating the integration of components implemented by crowd workers. We are considering potential approaches for managing progress and risk. \par

\begin{acks}
We thank the participants in the study for their participation. We also are grateful to Motoi Yamane, Masayuki Oda, Keiji
Kataoka, and Motoi Yamane at Piecemeal Technology for their assistance
in the case study. This work was supported in part by the National
Science Foundation under grants CCF-1414197 and CCF-1845508.

\end{acks}

\bibliographystyle{ACM-Reference-Format}
\bibliography{microtasking.bib}

\end{document}